\documentclass{PoS}

\newcommand{\re}{\mathbb{R}}

\title{$Sp(2n,\re)$ electric-magnetic duality as {\em off-shell} symmetry of interacting electromagnetic and scalar fields}

\ShortTitle{Electric-Magnetic Duality}

\author{Claudio Bunster\\
        Centro de Estudios Cient\'{\i}ficos (CECS), Casilla 1469, Valdivia, Chile\\
        E-mail: \email{bunster@cecs.cl}}

\author{\speaker{Marc Henneaux}\\
        Universit\'e Libre de Bruxelles and International Solvay Institutes, ULB-Campus Plaine CP231, B-1050 Brussels, Belgium; \\Centro de Estudios Cient\'{\i}ficos (CECS), Casilla 1469, Valdivia, Chile \\
       E-mail: \email{henneaux@ulb.ac.be}}

\abstract{It was established long ago that $SO(2)$ electric-magnetic duality is an {\em off-shell} symmetry of the free Maxwell theory, i.e., that it leaves invariant the action and not just the equations of motion.  We review here that analysis and extend it to the Maxwell field coupled to scalar fields defined on the $SL(2,\re)/SO(2)$ coset space, showing that $SL(2,\re)$ is in that case an {\em off-shell} symmetry. We also show how the result can be generalized to many Maxwell fields and $Sp(2n, \re)$ duality symmetry - or a subgroup of it,  recovering in particular the case of maximal supergravity in four dimensions with $E_{7,7}$ symmetry. We finally indicate further possible extensions to twisted self-duality equations for $p$-forms, including Chern-Simons terms and Pauli couplings,   as well as linearized gravity, which will be treated in depth elsewhere.}

\FullConference{Quarks, Strings and the Cosmos - H\'ector Rubinstein Memorial Symposium\\
		August 09-11, 2010\\
		(AlbaNova, Stockholm) Sweden}

\begin{document}

\section{Introduction}

\setcounter{equation}{0}
\subsection{$SO(2)$-Duality}

The vacuum Maxwell equations 
\begin{equation}
\partial_\mu \, F^{\mu \nu} = 0; \; \; \; \partial_\mu \, ^*\!F^{\mu \nu} = 0
\end{equation}
are invariant under electric-magnetic duality transformations, i.e.,  internal rotations in the two-dimensional plane of the electric and magnetic fields.  In covariant form, these transformations read
 \begin{eqnarray}
F^{\mu \nu} \; \; &\rightarrow& \; \;  \cos \alpha \, F^{\mu \nu} - \sin \alpha \;  ^*\!F^{\mu \nu} \label{DualF}\\
 ^*\!F^{\mu \nu} \; \; &\rightarrow& \; \;  \sin \alpha \,  F^{\mu \nu} + \cos \alpha \; ^*\!F^{\mu \nu}. \label{DualF*}
\end{eqnarray} 
Here, the dual field $^*\!F^{\mu \nu}$ is defined through\footnote{We use the mostly $+$'s signature for the Minkoswki metric and $\epsilon_{0123} = 1 = - \epsilon^{0123}$, $\epsilon^{123} = 1$.}
\begin{equation}
^*\!F^{\mu \nu} =  \frac{1}{2} \epsilon^{\mu \nu \rho \sigma} F_{\rho \sigma}
\end{equation}

\subsection{Invariance of the Maxwell Action}

Although it is often incorrectly stated that electric-magnetic duality  is only an ``on-shell" symmetry, i.e.,  only a symmetry of the equations of motion and not of the action, it was shown in \cite{Deser:1976iy} that this transformation is in fact a genuine symmetry transformation that leaves  the action invariant or, as one says in brief, is an ``off-shell symmetry".   The fact that duality is an off-shell symmetry is quite important since it permits its discussion at the quantum level through the path integral and also, it enables one, already at the classical level, to use the Noether method. 

In order to establish the duality invariance of the Maxwell action, 
\begin{equation} S[A_\mu] = - \frac{1}{4} \int d^4x F_{\mu \nu} F^{\mu \nu} 
\end{equation}
one must first extend ``off-shell" the duality transformations and express them in terms the dynamical variables, i.e., the components of the potential.  One appropriate extension\footnote{The explicit check of the invariance of the Maxwell action under duality transformations is done in \cite{Deser:1982xxyy} following \cite{Deser:1976iy}.  Our formulas are slightly different from those of \cite{Deser:1982xxyy} in that we keep $A_0$ and do not asume that the electric field is transverse off-shell.} is 
\begin{equation} \delta A_0 = 0 , \; \; \; \;  \delta A_k =   \beta \epsilon_{kpq}\, \triangle^{-1} (\partial^p E^{q}) \label{DualA}
\end{equation}
for infinitesimal duality rotations of parameter $\beta$.  Here, $E^k$ is the electric field, while $B^k$ below is the magnetic field,
\begin{equation}
E^k = - F^{0k}, \; \; \; B^k = \frac{1}{2} \epsilon^{kmn} F_{mn}.
\end{equation}
The transformation (\ref{DualA}) implies
\begin{equation} \delta E^k = - \beta B^k + \beta \epsilon^{kpq} \triangle^{-1} (\partial_p \partial^\mu F_{\mu q}) \label{DualE} \end{equation}
and
\begin{equation}
\delta B^k = \beta E^k -\beta \triangle^{-1}(\partial^k \partial_m E^m) \label{DualB}
\end{equation}  Hence, the infinitesimal transformations (\ref{DualE}) and (\ref{DualB}) derived from (\ref{DualA}) coincide on-shell with (the infinitesimal form of) (\ref{DualF}) and (\ref{DualF*}).

To verify that the Maxwell action is invariant under the duality transformations (\ref{DualA}), one checks that the variations of the kinetic term $(-1/2) F_{0k} F^{0k}$ and the potential term $(-1/4) F_{mn} F^{mn}$ are separately equal to total derivatives.  We start with the potential term.  One has
$$\frac{1}{4} \delta \left( F_{mn} F^{mn} \right) = B_k \delta B^k = B_k \left( \beta E^k -\beta \triangle^{-1}(\partial^k \partial_m E^m) \right).$$  The first term is proportional to the characteristic class $\epsilon^{\lambda \mu \rho \sigma} F_{\lambda \mu} F_{\rho \sigma}$ and thus it is a total derivative,
$$ B_k E^k = \frac{1}{2} \epsilon^{kmn} F_{mn} F_{0k} = \epsilon^{kmn} \partial_m A_n \partial_0 A_k + \frac{1}{2} \partial_k \left( \epsilon^{kmn} F_{mn} A_0 \right)$$
whereas \begin{equation}  \epsilon^{kmn} \partial_m A_n \partial_0 A_k = \partial_m\left(\epsilon^{kmn} A_n \partial_0 A_k\right) + \partial_0 \left(\epsilon^{kmn} \partial_m A_n A_k\right). \label{characteristic} \end{equation} The second term is also a total derivative since $B_k$ is divergence-free,
$$B_k  \left(  \triangle^{-1}(\partial^k \partial_m E^m) \right) =  \partial^k \left( B_k \triangle^{-1}(\partial_m E^m) \right).$$ The proof of the invariance of the kinetic term proceeds similarly. One finds explicitly
$$ \frac{1}{2}\delta \left(F_{0k} F^{0k} \right) = - \epsilon^{kmn}  E_k \triangle^{-1} (\partial_0 \partial_m E_n) = -\partial_0 \left( \epsilon^{kmn}  E_k \triangle^{-1} ( \partial_m E_n)\right) +  \epsilon^{kmn}  \triangle^{-1} ( \partial_m E_n)  \partial_0 E_k.$$  The last term has the same form as (\ref{characteristic}) with $A_k$ replaced by $E_k$ and the symmetric operator $\triangle^{-1}$ inserted and is thus also a total derivative.

As it was already shown in  \cite{Deser:1976iy},  the invariance of the action is easily verified to remain true when the coupling to gravity is switched on.  

\subsection{Manifestly duality-invariant action}

It would seem fair to state that, in the discussion given above, the invariance of the Maxwell action cannot be said to be manifest.  This is in part due to the fact that the electric-magnetic duality transformations are non local in space (but local in time) when expressed in terms of the vector potential $A_\mu$.  These somewhat awkward features can be remedied by following the procedure devised in \cite{Deser:1976iy}.  First, one goes to the Hamiltonian formalism (which was actually the starting point of  \cite{Deser:1976iy}), introducing the momenta $\pi^i$ conjugate to $A_i$ as independent variables in the variational principle.  These momenta coincide in the case of the pure Maxwell theory with the electric field $E^i$. Second, one solves Gauss'law for $\pi^i$, introducing a second vector potential $Z_i$.  This step gets rid at the same time of the time component $A_0$.  In terms of the two spatial vector potential ${\mathbf A}_a = ( {\mathbf A}, {\mathbf Z})$, $a=1,2$, the action reads  \cite{Deser:1976iy,Deser:1997mz}
\begin{equation}
S^{\hbox{{\tiny inv}}}[A^a_i] = \frac{1}{2}  \int dx^0 \, d^3x \left( \epsilon_{ab} {\mathbf B}^a \cdot \dot{{\mathbf A}}^b  - \delta_{ab} {\mathbf B}^a \cdot {\mathbf B}^b \right), \; \; a, b = 1,2 \label{Action0}
\end{equation}
where 
$$ {\mathbf B}^a  = {\mathbf \nabla} \times {\mathbf A}^a$$ and $\epsilon_{ab} = - \epsilon_{ba}$ is the Levi-Civita tensor in 2 dimensions (with $\epsilon_{12}  = 1$).
The duality rotations in this formulation are
\begin{eqnarray}
{\mathbf A}^1 \; \; &\rightarrow& \; \;  \cos \alpha \, {\mathbf A}^1 - \sin \alpha \,  {\mathbf A}^2 \\
{\mathbf A}^2 \; \; &\rightarrow& \; \;  \sin \alpha \,  {\mathbf A}^1 + \cos \alpha \, {\mathbf A}^2.
\end{eqnarray}
The action (\ref{Action0}) is clearly invariant under these transformations since both $\epsilon_{ab}$ and $\delta_{ab}$ are invariant tensors for $SO(2)$. The standard magnetic field $ {\mathbf B}$  is $  {\mathbf B} \equiv {\mathbf B}^1$ while the standard electric field $ {\mathbf E}$ is  $ {\mathbf E} \equiv - {\mathbf B}^2$. Although manifestly duality invariant, this reformulation of the Maxwell action is not manifestly Lorentz invariant.

It is important to realize that (\ref{Action0}) is a mere rewriting of the standard Maxwell action.  The classical steps that go from one form of the Maxwell action to the other can be repeated quantum-mechanically in the path integral, as either insertion of delta functions (for including the constraints) or Gaussian integration over the momenta (to go from the first-order to the second-order form of the action).  In our case, the accompanying determinant factors, which appear in the path integral measure in the second-order formalism,  are just $c$-numbers.  

The possibility of introducing a second potential in order to make the electric-magnetic duality manifest stems technically from the fact that the gauge constraints can be written as a divergence.  It is quite remarkable, and by no means obvious, that this can also be achieved in the case of linearized gravity \cite{Henneaux:2004jw}.  There, had it not been for the insistence in raising electric-magnetic duality invariance to the majesty of a principle that should have a manifest expression, one would hardly have been led to discover those potentials.  As we shall see below, this point of view can be successfully implemented also in {\em interacting} theories, which gives additional weight to its adoption as a sound physical principle.

\subsection{$U(n)$-Duality}
\label{2.2}
{}For  $n$  Maxwell fields, the manifestly duality invariant action takes the form (\ref{Action0}) but the internal indices run now from $1$ to $2n$ values and the $\epsilon$-symbol is replaced by the antisymmetric canonical symplectic form $\sigma_{MN}$ ($M, N = 1,\cdots, 2n$),
$$ \sigma = \left(\begin{array}{ccccccc} 0 & 1 & 0 & 0 & \cdots & 0 &0 \\ -1 & 0 & 0 & 0 & \cdots & 0&0\\ 0 & 0 & 0 & 1 & \cdots & 0&0\\ 0 & 0 & -1& 0& \cdots & 0&0 \\ \vdots&\vdots&\vdots& \vdots&\vdots & \vdots & \vdots\\ 0&0&0&0 &\cdots & 0 & 1 \\ 0&0&0&0& \cdots &-1 & 0\end{array} \right), $$
which gives
\begin{equation}
S^{\hbox{{\tiny inv}}}[A^M_i] = \frac{1}{2}  \int dx^0 \, d^3x \left( \sigma_{MN} {\mathbf B}^M \cdot \dot{{\mathbf A}}^N  - \delta_{MN} {\mathbf B}^M \cdot {\mathbf B}^N \right). \label{Action1}
\end{equation}

The duality group clearly contains in this case $n$ factors $SO(2)\times SO(2) \times \cdots \times SO(2)$, namely, one $SO(2)$ for each pair $(A^{2k-1},A^{2k})$ describing a single standard Maxwell field. But duality is in fact bigger, because one can also perform linear transformations that mix the vector potentials belonging to different pairs. Hence  the duality group is enlarged from $[SO(2)]^n$ to $U(n)$ \cite{Ferrara:1976iq,Gaillard:1981rj,deWit:2001pz,Bunster:2010wv}.   This can be seen as follows: linear transformations of the potentials $A^M$,
\begin{equation} 
A^M \rightarrow A'^M = \Lambda^M_{\; \; N} A^N
\end{equation}
where $\Lambda \in GL(2n,\re)$, leave the action (\ref{Action1}) invariant if and only if they preserve the symplectic product and the scalar product (in order to preserve the kinetic term and the Hamiltonian, respectively),
\begin{equation} 
\Lambda^T \sigma \Lambda = \sigma, \; \; \; \Lambda^T I \Lambda = I.
\end{equation}
The first condition implies $\Lambda \in Sp(2n, \re)$, while the second implies $\Lambda \in O(2n)$ ($O(2n)$ always means here $O(2n, \re)$).  Accordingly, the transformation $\Lambda$ must belong to $Sp(2n,\re) \cap O(2n)$, which is the maximal compact subgroup of the symplectic group $Sp(2n,\re)$, known to be isomorphic to $U(n)$ (see, for example, \cite{Helgason}).

In infinitesimal form, the invariance condition reads, with $\Lambda = I + \lambda$,
\begin{equation} 
\lambda^T \sigma + \sigma \lambda = 0, \; \; \; \lambda^T + \lambda = 0,
\end{equation}
or in component form,
\begin{equation} 
 \sigma_{PN} \lambda^P_{\; \; M}  +\sigma_{MP} \lambda^P_{\; \; N} = 0, \; \; \; \delta_{PN} \lambda^P_{\; \; M}  +\delta_{MP} \lambda^P_{\; \; N} = 0. \label{InvarianceCond}
\end{equation}
 
By reversing the steps that lead from the second order formalism to the first order formalism, one can rewrite the action (\ref{Action1}) in second order form.  It is just the sum of $n$ Maxwell actions. The first order and second order forms of the action share the same symmetries, in particular, the same duality symmetries, although some of these take a non-local form in the second order formalism (see \cite{Henneaux:1992ig} for general information on symmetries in the first and second order formulations, which are related by changes of variables and introduction (or elimination) of auxiliary fields). So, contrary to the folklore,  the sum of $n$ Maxwell actions is invariant under the full $U(n)$ duality group, even though it is only manifestly so under $SO(n)$ rotations in the space of the $n$ spacetime vector potentials. 

\subsection{$Sp(2n,\re)$-duality}
It has been realized in the context of supergravity that the compact $U(n)$ duality invariance can be extended to a non-compact $Sp(2n,\re)$ invariance (or a subgroup of it) if couplings to appropriate scalar fields are introduced \cite{Cremmer:1977tt,Cremmer:1978ds,Cremmer:1979up,Gaillard:1981rj,Breitenlohner:1987dg}.  For $Sp(2n,\re)$ invariance, the scalar fields parametrize the coset space $Sp(2n, \re) /U(n)$. Invariance was established only at the level of the equations of motion.  The purpose  of this paper is to show that this extended duality invariance is again, in fact, a proper {\em off-shell} invariance of the action, contrary to somewhat fatalistic widespread fears. We start with the simple case, $n=1$.  We then consider the general case.

\section{The case $n=1$}

\subsection{Second order Lagrangian}
{}For $n=1$, $Sp(2,\re) \simeq SL(2, \re)$, $U(1) \simeq SO(2)$ and the appropriate scalar fields come in a pair $(\phi, \chi)$ parametrizing the coset space $SL(2, \re) /SO(2)$.  The scalar field $\phi$ is the ``dilaton" while $\chi$ is the ``axion".  The Lagrangian for the scalars is given by
\begin{equation}
L^S = - \frac{1}{2} \partial_\mu \phi \partial^\mu \phi - \frac{1}{2} e^{2\phi} \partial_\mu \chi \partial^\mu \chi
\end {equation}
and is invariant under $SL(2,\re)$ transformations, which are non linear and read in infinitesimal form
\begin{equation}
\delta \phi =  \epsilon^\alpha \, \xi_\alpha^\phi(\phi,\chi), \;Ê\; \; \; \delta \chi = \epsilon^\alpha \, \xi_\alpha^\chi(\phi,\chi), \; \; \; \; \alpha = +,0,-. \label{TransScalars}
\end{equation}
Here the Killing vector fields $\xi_\alpha$ tangent to the coset space are explicitly given by
\begin{eqnarray}
&&\xi_- = \frac{\partial}{\partial \chi} \\
&& \xi_0 = 2 \frac{\partial}{\partial \phi} - 2 \chi \frac{\partial}{\partial \chi} \\
&& \xi_+ =  2 \chi \frac{\partial}{\partial \phi} + \left(e^{-2 \phi} - \chi^2 \right) \frac{\partial}{\partial \chi}
\end{eqnarray}
and fulfill the $SL(2,\re)$ algebra in the Lie bracket, $[\xi_+, \xi_0] = 2 \xi_+$, $[\xi_-, \xi_0] = -2 \xi_-$, $[\xi_-, \xi_+] =  \xi_0$.
The rigid $SO(2)$ transformations are generated by $\xi_+ - \xi_-$.  This is the stability subgroup of the ``origin" $(\phi, \chi) = (0,0)$ since there, the $SL(2,\re)$ transformations reduce to $$\delta \phi = 2 \epsilon^0, \; \; \; \delta \chi = \epsilon^+ + \epsilon^-.$$

Turn now to the Lagrangian for the vector field $A_\mu$.  It was shown in \cite{Gaillard:1981rj,Breitenlohner:1987dg} that $SL(2,\re)$ invariance of the equations of motion for $A_\mu$ is achieved provided the couplings of the vector field to the scalar fields are chosen as
\begin{equation}
L^V = - \frac{1}{4} e^{-\phi} F_{\mu \nu} F^{\mu \nu} + \frac{1}{8} \chi \epsilon^{\lambda \mu \rho \sigma} F_{\lambda \mu} F_{\rho \sigma}
\end{equation}
The action for the coupled system is
\begin{equation}
S = \int d^4 x (L^V + L^S)
\end{equation}
and leads to equations of motion that are $SL(2,\re)$-invariant.

The goal is to prove that the action {\em is also invariant} under $SL(2, \re)$-duality transformations, which are therefore off-shell symmetries.  This extends to the coupled system the analysis made above for the pure Maxwell case, recovered by choosing a particular value of the scalar fields, say $(\phi, \chi) = (0,0)$. As we have already indicated, the $SL(2,\re)$ transformations that preserve these values form a $SO(2)$ subgroup.

\subsection{First order form of the vector Lagrangian}
To prove invariance, it suffices to prove that $\int d^4x L^V$ is invariant since the scalar term is manifestly so.  To that end, we rewrite $\int d^4x L^V$ in a form where this invariance is manifest, by going to the first order formalism and solving Gauss' law. As we shall see, the resulting first-order action is indeed manifestly $SL(2,\re)$-invariant.  This invariance lifts to the original second order form of the action, just as in the pure Maxwell case, and for the same reasons indicated in that case.

\subsubsection{Hamiltonian form of the vector action}
The  momenta $\pi^i$ conjugate to $A_i$ are
\begin{equation}
\pi^i = - e^{-\phi} F^{0i} + \frac{1}{2} \chi F_{jk}\epsilon^{ijk}
\end{equation}
from which one derives the Hamiltonian form of the vector action
\begin{equation}
S^{V,H} = \int d^4x \left(\pi^i \dot{A}_i - {\mathcal H} - A_0 {\mathcal G} \right)
\end{equation}
where
\begin{equation}
{\mathcal H} = \frac{1}{2} \left(e^\phi \pi^i \pi_i + \chi e^\phi \epsilon^{imn} \pi_i F_{mn} + \frac{1}{2}(\chi^2 e^\phi + e^{-\phi})F^{mn}F_{mn}\right)
\end{equation}
and
\begin{equation}
{\mathcal G} = - \partial_i \pi^i .  \label{Gauss}
\end{equation}
So we find again the fundamental feature that the gauge constraint is a divergence.

\subsubsection{Solving Gauss' law}
Because Gauss' law takes exactly the same form as in the absence of scalar fields, it can be solved in exactly the same way, by introducing a second vector potential $Z_i$, \begin{equation}
\pi = -\frac{1}{2}  \epsilon^{imn} H_{mn}, \; \; \; \; H_{mn} = \partial_m Z_n - \partial_n Z_m .  
\end{equation}
As stressed above, this is the key to exhibit manifest duality invariance.  When doing so, the Lagrange multiplier $A_0$ drops out and the vector action takes the form
\begin{equation}
S^{V,\hbox{{\tiny inv}}}[A^a] = \frac{1}{2}  \int dx^0 \, d^3x \left( \epsilon_{ab} {\mathbf B}^a \cdot \dot{{\mathbf A}}^b  - G_{ab}(\phi, \chi) {\mathbf B}^a \cdot {\mathbf B}^b \right), \; \; a, b = 1,2 \label{ActionV0}
\end{equation}
with
\begin{equation} (G_{ab}) = \left(\begin{array}{cc} \chi^2 e^\phi + e^{-\phi}\; \; \; & - \chi e^\phi  \\  -\chi e^\phi&e^\phi\end{array} \right). 
\end{equation}
This is exactly the same expression as (\ref{Action0}), but with the Euclidean metric $\delta_{ab}$ replaced by the scalar field dependent metric $G_{ab}$, to which $G_{ab}$ reduces at the origin of scalar field space.  Note that $G_{ab}$ has Euclidean signature and determinant equal to $+1$.

\subsection{Manifest $SL(2,\re)$-invariance}
The replacement of the Euclidean metric by $G_{ab}$ enlarges the symmetry from $SO(2)$ to $SL(2,\re)$.  This is because under a general infinitesimal $SL(2,\re)$ transformation, 
\begin{equation}
\delta {\mathbf A}^a = \epsilon^\alpha \left(X_\alpha\right)^a_{\; \; b} {\mathbf A}^b; \; \; \; \; \delta {\mathbf B}^a = \epsilon^\alpha \left(X_\alpha\right)^a_{\; \; b} {\mathbf B}^b \label{TransElec}
\end{equation}
under which the scalar fields transform as in (\ref{TransScalars}), one finds
\begin{equation}
\delta G_{ab} + \epsilon^\alpha G_{cb} \, \left(X_\alpha\right)^c_{\; \; a} + \epsilon^\alpha G_{ac} \, \left(X_\alpha\right)^c_{\; \; b} = 0
\end{equation}
so that the Hamiltonian is invariant.  Here, $\delta G_{ab} = \frac{\partial G_{ab}}{\partial \phi} \delta \phi + \frac{\partial G_{ab}}{\partial \chi} \delta \chi $ while the three matrices $X_\alpha$ are the generators of $SL(2,\re)$,
\begin{equation} X_- = \left(\begin{array}{cc} 0& 0  \\  1&0 \end{array} \right), \; \; \;  X_0 = \left(\begin{array}{cc} 1& 0  \\  0&-1 \end{array} \right), \; \; \; X_+ = \left(\begin{array}{cc} 0& 1  \\  0&0 \end{array} \right).
\end{equation}
Similarly, the kinetic term is invariant under (\ref{TransElec}), already without the scalar fields, since the symplectic form $\epsilon_{ab}$ is invariant under $SL(2,\re) \simeq Sp(2,\re)$ transformations.  

It follows that the vector action is invariant under $SL(2,\re)$ transformations.  Accordingly, so is also the total action, in either first order or second order form.  $SL(2,\re)$ electric-magnetic duality  is an off-shell symmetry.

\section{The general case}
The extension to the general case of $n$ electromagnetic fields is straihtforward.  The manifestly invariant form of the vector action reads as in (\ref{Action1}) but with the Euclidean metric $\delta_{MN}$ replaced by a scalar field dependent metric $G_{MN}(\varphi^i)$
\begin{equation}
S^{\hbox{{\tiny inv}}}[A^M_i] = \frac{1}{2}  \int dx^0 \, d^3x \left( \sigma_{MN} {\mathbf B}^M \cdot \dot{{\mathbf A}}^N  - G_{MN} (\varphi^i){\mathbf B}^M \cdot {\mathbf B}^N \right), \; \; \; \; M, N = 1, \cdots, 2n. \label{ActionV1}
\end{equation}
The kinetic term in the action is invariant under $Sp(2n, \re)$ transformations,
\begin{equation}
\delta {\mathbf A}^M = \lambda^M_{\; \; N} {\mathbf A}^N; \; \; \; \; \delta {\mathbf B}^M = \lambda^M_{\; \; N} {\mathbf B}^N \label{TransElec1}
\end{equation}
with 
\begin{equation}
\lambda^T \sigma + \sigma \lambda = 0.
\end{equation}
The coupling to the scalar fields does not affect this property, since the scalar fields do not enter the kinetic term.  The invariance under $Sp(2n,\re)$ will extend to the Hamiltonian if the variation $\delta G_{MN}$ of the metric due to the variation of the scalar fields compensates the variation of the magnetic fields, i.e.,
\begin{equation}
\delta G_{MN} + G_{QN} \lambda^Q_{\; \; M} + G_{MQ} \lambda^Q_{\; \; N} = 0  \; \; \; \; \Leftrightarrow \; \; \; \; \delta G + \lambda^T G + G \lambda = 0 \label{keycrucial}
\end{equation}
This condition replaces the more restrictive condition $\lambda^T + \lambda = 0$ which must be imposed on $\lambda$ in the absence of scalar fields. 

One way to achieve (\ref{keycrucial}) is to take the scalar fields in the coset space $Sp(2n,\re) /U(n)$ and to construct $G_{MN}$ from (\ref{keycrucial}), which is then viewed as a set of equations that determine $G_{MN}$ rather than conditions on $\lambda \in Sp(2n, \re)$.  The solution for $G_{MN}$ is given for instance in  \cite{Breitenlohner:1987dg}.

Once the first-order action is completely known, one can go to its second order form by (i) trading $n$ of the potentials (conventionally the even-numbered ones) for the momenta $\pi^{a \, i}$ conjugate to the $n$ potentials $A^a_i$ ($a= 1, \cdots, n$) that are kept, through the relation $\pi^{a \, i} = - \frac{1}{2} \epsilon^{imn} (\partial_m Z^a_n - \partial_n Z^a_n)$, and implementing the Gauss' constraints $-\partial_i\pi^{a \, i}   \approx 0$ through the Lagrange multipliers method, the Lagrange multipliers being the time components $A_0^a$; and (ii) integrating over the conjugate momenta to get the vector action $S^V[A^a_\mu]$ in second order form,
\begin{equation}
S^V[A^a_\mu] = - \frac{1}{4} \int d^4x \left(\mu_{ab}(\varphi^i) F^a_{\mu \nu} F^{b \, \mu \nu} + \nu_{ab}(\varphi^i) \epsilon^{\mu \nu \rho \sigma} F^a_{\mu \nu} F^b_{\rho \sigma} \right) \label{ActionVCov}
\end{equation}
to which one must add the action for the scalar fields $\varphi^i \in SL(2n, \re)/U(n)$.  Here, $\mu_{ab}(\varphi^i)$ and $\nu_{ab}(\varphi^i)$ are functions of the scalar fields $\varphi^i$ determined from $G_{MN}(\varphi^i)$ through the process of going from the first order form to the second order form of the vector action.  These functions are given in \cite{Gaillard:1981rj,Breitenlohner:1987dg}, where they are obtained through the requirement that the equations of motion are invariant under $SL(2n,\re)$ (on-shell symmetry).  
But $SL(2,\re)$ electric-magnetic duality invariance is in fact a bona fide off-shell invariance as our analysis shows, although in the second order formalism, the transformations of the vector potentials are non-local in space and somewhat awkward.  

Conversely, if one knows the second order form (\ref{ActionVCov}) of the action, one can derive the first-order form (\ref{ActionV1}) by first introducing the momenta $\pi^i_a$ conjugate to $A^a_i$, which are found to be subject to Gauss'  constraints, and then solving explicitly the constraints through the introduction of the second set of potentials $Z^a_i$.

The group $Sp(2n,\re)$ is a maximal symmetry of the theory, implemented when the scalar fields belong to the coset $Sp(2n, \re)/U(n)$.  If one takes a smaller set of scalar fields, belonging to the coset $G/H$ where $G$ is a subgroup of $Sp(2n, \re)$ and repeats the above construction, one gets invariance under  a smaller group $G \subset Sp(2n, \re)$ of dualities. This is the situation for maximal supergravity in four dimensions where the duality group is $E_{7,7} \subset Sp(56,\re)$ \cite{Cremmer:1978ds,Cremmer:1979up}.  The first-order form of the action (\ref{ActionV1}) is given in \cite{Hillmann:2009zf} and extensively used in \cite{Bossard:2010dq}.  The group of dualities cannot be bigger than  $Sp(2n,\re)$ because the kinetic term does not (and cannot) depend on the scalar fields\footnote{A scalar field dependence of the kinetic term would clash with the gauge invariances of the vector fields.}.

\section{Conclusions}
\label{Conclusions}

In this article, we have explicitly shown that the $SL(2n,\re)$ duality of interacting electromagnetic and scalar fields  \cite{Cremmer:1977tt,Cremmer:1978ds,Cremmer:1979up,Gaillard:1981rj,Breitenlohner:1987dg} is, contrary to the existing folklore, an off-shell symmetry and not just a symmetry of the equations of motion.  This property extends the analysis of \cite{Deser:1976iy} performed for one free Maxwell field. Crucial in the existence of the duality symmetry is the fact that Gauss' law takes the form of a total divergence.  This property is lost when minimal coupling is considered, but is preserved by Pauli couplings, present in supergravity, as well as Chapline-Manton couplings or Chern-Simons terms, all of which  therefore preserve the symmetry. 

A similar analysis applies to (twisted) self-duality equations for $p$-forms \cite{Cremmer:1998px}, which also derive from a (non manifestly spacetime covariant) variational principle contrary to widespread fatalism.  A special case was already  treated in \cite{Henneaux:1988gg}.  We shall return to this question elsewhere  \cite{BH2011}.  Gravitational equations in diverse dimensions viewed as twisted self-duality equations will also be considered along the lines of \cite{Henneaux:2004jw}.

{}Finally, a word about the gauging of the $Sp(2n, \re)$-symmetry.  It was shown in \cite{Bunster:2010wv} and confirmed in \cite{Deser:2010it} that one cannot gauge the electric-magnetic duality symmetry in the case of free Maxwell fields by following the standard pattern of replacing abelian curvatures by non-abelian ones.  A comparison with earlier approaches may be found in \cite{Saa:2011wb}.  Since the obstruction to gauging comes from the impossibility of  embedding  the adjoint group (of any group) in $Sp(2n,\re)$  \cite{Bunster:2010wv}, and since the kinetic term responsible for this problem is unchanged when scalar couplings of the above type are included, this negative result about impossibility of gaugings {\em \`a la Yang-Mills} remains valid in that case.

\section*{Acknowledgments} M. H.  gratefully acknowledges support from the Alexander von Humboldt Foundation through a Humboldt Research Award and support from the ERC through the ``SyDuGraM" Advanced Grant.  The Centro de Estudios Cient\'{\i}ficos (CECS) is funded by the Chilean Government through the Centers of Excellence Base Financing Program of Conicyt. The work of M. H. is also partially supported by IISN - Belgium (conventions 4.4511.06 and 4.4514.08), by the Belgian Federal Science Policy Office through the Interuniversity Attraction Pole P6/11 and by the ``Communaut\'e Fran\c{c}aise de Belgique" through the ARC program.

\end{document}